\documentclass[a4paper,notitlepage]{article}
\usepackage{graphicx}
\usepackage{amsmath}
\usepackage{a4}
\usepackage{amsfonts}
\usepackage{amssymb}
\begin{document}

\title{Rayleigh-Taylor Instabilities in Thin Films of Tapped Powder}
\author{J. Duran\\LMDH- UMR\ 7603\ CNRS- Universit\'{e} P.\ et M.\ Curie\\4 place Jussieu, 75252 Paris Cedex 05\\email address : jd@ccr.jussieu.fr}
\maketitle
\begin{abstract}
We observe powder ''droplets'' forming when tapping repeatidly an horizontal
flat plate initially covered with a monolayer of fine powder particles.
Starting from a simple model involving both the air flow through the porous
cake and avalanche properties, we setup an analytical model which
satisfactorily fits the experimental results.\ We observe a close analogy
between the governing equations of the phenomenon and the basic physics of
wetting liquids, including the equivalent of the Laplace law and the surface
tension parameter leading to the well known Rayleigh Taylor instability. \ 
\end{abstract}


\bigskip\medskip

In the recent years, there has been a great deal of interest in the response
of granular materials\ to various kinds of external perturbations.\ Up to now,
the vast majority of the experimental, theoretical and simulated works have
dealt with model granular solids in the sense\cite{brown70} that the particles
were supposed to be large enough (i.e.\ typically larger than $100\mu m)$ to
avoid significant interaction with the surrounding fluids\cite{pak95}. In
reverse and rather paradoxically, the understanding of the behavior of fine
powders has received much less attention although it is universally recognized
as the keystone of an increasing number of high-tech industrial processes. In
this spirit, a few recent attempts were made towards the analysis of the
behavior of fine (typically in the range from $1$ to less than $30\mu m)$
cohesive or non-cohesive powders in air (e.g. \cite{raafat96}%
\cite{castellanos99}\cite{falcon99}\cite{laroche89}\cite{duran00}) or of
larger particles submitted to excessive windy conditions\cite{pak95}%
\cite{tsimring99}, such as in desert dunes\cite{kroy01}.

In a recent paper\cite{duran00} we reported a series of experiments and an
analytical model dealing with a thick (about 10mm) slice of fine powder in a
container tapped from below.\ We reported that, under these circumstances, a
quasi periodic corrugated pattern spreads out.\ The characteristic wavelength
of this pattern was found to be proportional to the amplitude of the taps. Our
theoretical model involved two basic features of the fine granular material :
First, the Darcy's law for modelling the air flux trapped under the layer and
pushed through the porous cake of granulate when the pile falls down
immediately after the shock.\ Secondly, it involved the maximum stability
angle of a granulate before generating avalanches.\ It was shown that a
corrugated surface stands as a more stable state than a horizontal flat
surface with respect to air blow from below, because it is easier to eject a
particle from a flat surface than from an inclined surface sitting at the
avalanche angle. We explained that the apices of the hills created by air blow
from below are unstable as compared to both sides of the hills. A couple of
further experiments showing direct ''crater like'' formation resulting from
this ''volcano effect'', provided us a clear view of this last feature.%

%TCIMACRO{\FRAME{fhFU}{2.4811in}{1.7573in}{0pt}{\Qcb{Sketch of the trajectories
%of the powder particles participating to the intrinsic convection process when
%the heap are ejected above the plate resulting from either taps or air blowing
%from below.}}{\Qlb{fig1}}{fig1.eps}{\special{ language "Scientific Word";
%type "GRAPHIC";  maintain-aspect-ratio TRUE;  display "USEDEF";
%valid_file "F";  width 2.4811in;  height 1.7573in;  depth 0pt;
%original-width 7.7824in;  original-height 5.4967in;  cropleft "0";
%filename 'fig1.eps';file-properties "XNPEU";}}}%
%BeginExpansion
\begin{figure}
[h]
\begin{center}
\includegraphics[
height=1.7573in,
width=2.4811in
]%
{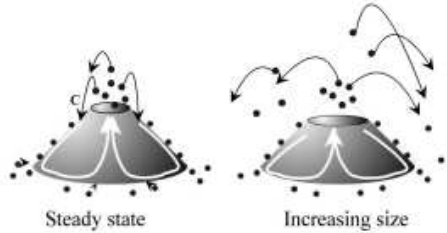}%
\caption{Sketch of the trajectories of the powder particles participating to
the intrinsic convection process when the heap are ejected above the plate
resulting from either taps or air blowing from below.}%
\label{fig1}%
\end{center}
\end{figure}
%EndExpansion

Thus, our model explicitly accounts for the earlier Faraday's
observation\cite{faraday31} of the ''chimney or volcano effect''.\ Faraday
stated that '' It forms a partial vacuum into which the air round the heap
enters with more readiness than the heap itself. The particles of the heap
rise up at the center, overflow, fall down upon all slides and disappear at
the bottom, apparently proceeding inwards'' as sketched in Fig.\ref{fig1}
which, additionally shows the resulting inner convections rolls.

Keeping along the same line, we make a step further considering now a thin
slice of a fine powder (typical particle size $30%
%TCIMACRO{\unit{\UNICODE{0x3bc}m}}%
%BeginExpansion
\operatorname{\mu m}%
%EndExpansion
)$ spread out over a flat plate. When gently tapping repeatedly and at
constant intensity onto the plate, we observe the formation of a collection of
separate rounded conical heaps looking like droplets of powder evenly spread
over the plate.\ The resulting pattern strikingly reminds one of the
Rayleigh-Taylor instability illustrated by the droplets structure obtained
when turning up a glass plate initially covered with a thin film of a wetting
liquid. As we will show in the following, this analogy is not fortuitous.\ It
results from an underlying similarity between the equations governing the
wetting properties of liquids and the behavior of powder piles interacting
with a surrounding gas.

\begin{figure}
[h]
\begin{center}
\includegraphics[
height=2.4855in,
width=2.6074in
]%
{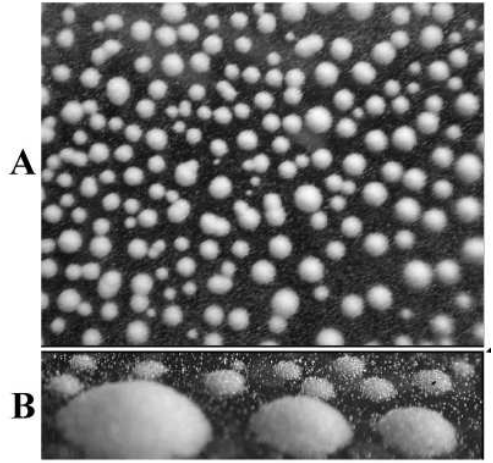}%
\caption{Above : Bird eye view of the pattern obtained after fourty taps over
a plate initially covered with an approximately uniform film of powder
particles (dia. about 30 microns).\ The mean distance between neighbouring
heaps is about 5mm.\ Below : The snapshot shows an oblique view of a few small
heaps.\ It exhibits the rounded shape of the apices due to the ''volcano
effect''.}%
\label{fig2}%
\end{center}
\end{figure}

Several basic characteristic features of the instability of a tapped thin film
of powder can be readily observed starting from a simple table-top experiment
: Using a small leucite rule equipped with thin spacers, we spread a monolayer
slice of powder (silica beads, diameter about 30$\mu m$) over a flat glass
plate (size 6x9 $cm^{2})$.\ This glass plate is kept horizontal and secured on
its periphery using a latex band which allows a certain degree of freedom for
up and down motions. Using a small metallic or plastic rod, we knock gently
and repeatedly at a very low pace and at a constant intensity over one corner
of the glass plate, applying vertically as brief taps as possible. After a few
taps (about ten to twenty), the surface, initially flat, smooth and
horizontal, separates into a collection of tiny rounded conical heaps looking
like droplets similar to those reported in Fig.\ref{fig2}. Starting from the
same initial conditions but tapping more energetically while keeping the
intensity as constant as possible from one tap to the next, induces a pattern
with bigger droplets separated by a larger distance. Note that excessive
wetness prevents the observation of these surface patterns.%

%TCIMACRO{\FRAME{fhFU}{2.6074in}{2.4855in}{0pt}{\Qcb{Above : Bird eye view of
%the pattern obtained after fourty taps over a plate initially covered with an
%approximately uniform film of powder particles (dia. about 30 microns).\ The
%mean distance between neighbouring heaps is about 5mm.\ Below : The snapshot
%shows an oblique view of a few small heaps.\ It exhibits the rounded shape of
%the apices due to the ''volcano effect''.}}{\Qlb{fig2}}{fig2.eps}%
%width 2.6074in;  height 2.4855in;  depth 0pt;  original-width 11.336in;
%original-height 10.805in;  cropleft "0";  croptop "1";  cropright "1";
%cropbottom "0";  filename 'fig2.eps';file-properties "XNPEU";}}}%
%BeginExpansion
%EndExpansion

\begin{figure}
[h]
\begin{center}
\includegraphics[
height=2.3073in,
width=2.9637in
]%
{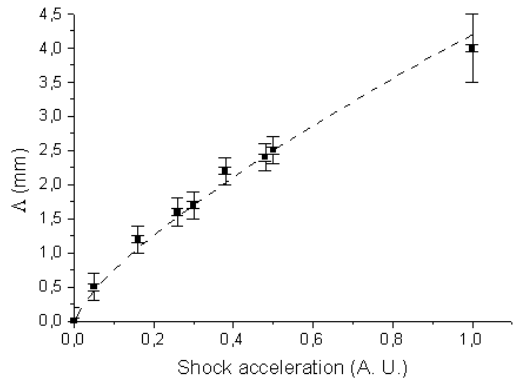}%
\caption{Experimental results obtained with a monolayer slice of silica powder
(particle size about 35 $\mu m).$The dashed line is a theoretical best fit to
Eq. (9)}%
\label{fig3}%
\end{center}
\end{figure}

Definitely more reliable information has been obtained in the course of our
experiments, using a more sophisticated device. We set a CCD (charge coupled
device) camera above the plate in order to record and process the successive
patterns obtained during the experiments.\ Secondly, we used a magnetically
driven tapping device and a commercial Bruer and Kjaer accelerometer stuck on
the plate in the vicinity of the sample in order to monitor the acceleration
of the taps. Thus we get the wavelength dependence on the tap acceleration.
Typical experimental results are reported in Fig.\ref{fig3}%

%TCIMACRO{\FRAME{fhFU}{2.9637in}{2.3073in}{0pt}{\Qcb{Experimental results
%obtained with a monolayer slice of silica powder (particle size about 35 $\mu
%m).$The dashed line is a theoretical best fit to Eq. (9)}}{\Qlb{fig3}%
%}{fig3.eps}{\special{ language "Scientific Word";  type "GRAPHIC";
%width 2.9637in;  height 2.3073in;  depth 0pt;  original-width 6.6945in;
%original-height 5.2027in;  cropleft "0";  croptop "1";  cropright "1";
%cropbottom "0";  filename 'fig3.eps';file-properties "XNPEU";}}}%
%BeginExpansion
%EndExpansion

First we look for a relationship between the height of the approximately
identical conical piles and the mean distance separating them.\ Consider the
initial situation when a thin slice of powder of thickness $e$ made of small
spherical beads (diameter $D)$ is evenly spread over a horizontal flat surface
whose area is $S.$ Suppose now that the powder has been gathered in a number
of $N$ disjointed identical conical piles having an angle $\theta$ to
horizontal and culminating at altitude $h.$ These piles are evenly distributed
over the area $S$. Due to volume conservation, the number $N$ of these piles
is approximately given by%

\begin{equation}
N=3\frac{Se\tan^{2}\theta}{\pi}\frac{1}{h^{3}}\propto h^{-3} \label{volume}%
\end{equation}

The wavelength $\Lambda$ of this pattern is the square root of the mean area
occupied by each pile%

\begin{equation}
\Lambda=\sqrt{\frac{\pi}{3e\tan^{2}\theta}}\quad h^{\frac{3}{2}}%
\propto\ h^{\frac{3}{2}} \label{lambda}%
\end{equation}

In agrement with Faraday's and more recent authors'\cite{laroche89}
observation of the volcano effect and the associated convection, we have
shown\cite{duran00} that when a conical powder pile undergoes a ballistic
flight, we can distinguish between two regions, delimitated by a circle at
altitude $h_{C}$ \ (Fig.\ref{fig1}). The lowest region is stable against the
upcoming air flux because it is stabilized by the lateral avalanches ($0\leq
h<h_{C})$.\ Around the apex we found an unstable part ($h_{C}\leq h\leq
h_{T})$ (T for top) where grains are expelled by the upcoming air flux. The
adimensional parameter $C$ measures the proportion of the unstable part of the
heap, so that $C=\left(  h_{T}-h_{C}\right)  /h_{T}.$ In our preceding paper
dealing with thick powder layers\cite{duran00}, we observed that the stability
of the corrugated patterns and the linear dependence of the wavelength on the
shocks amplitude are consistent with the fact that $C$ is independent of the
shocks acceleration.\ Stated differently, the steady state of a pattern as
sketched in Fig.\ \ref{fig1} results from the balance between the number of
expelled particles \ near the apices and the number of particles which are
reinjected into the bulk of the heaps at every tap.

In order to eject a single\ free particle of diameter $D$ and density $\rho,$
sitting on an horizontal surface, we need an air velocity which exceeds the
free flight velocity of this particle $v_{f}=D^{2}\rho g/18\eta$ where $\eta$
is the air viscosity and $g$ the gravitational acceleration. Now, we consider
a particle sitting at altitude $h_{C}$ on the inclined surface of the heaps$.$
We built a simplified equation\cite{duran00} for the screening effect due to
the avalanche process, considering that the mass of the particles lying at an
altitude $h$ is increased by a factor $Np\sin\theta,$ $N$\ being the number of
the above lying particles pertaining to a single sheet of the inclined
granulate and $p$ being the number (typically 5) of sheets possibly involved
in the avalanche. Thus the required air velocity needed to eject the
considered particle is given by $v_{h_{C}}=v_{f}\left(  h_{T}-h_{C}\right)
p\sin\theta/D.\;$

The velocity of the upcoming air flux at altitude $h_{C}$ is given \ by the
Darcy's law, so that :%

\begin{equation}
v_{h_{C}}=\frac{K\Delta P}{\eta h_{C}}=v_{f}\frac{C}{1-C}\frac{h_{C}%
p\sin\theta}{D}%
\end{equation}

where $K$ is the permeability of the powder and $\Delta P$ the pressure
difference acting over the granular porous cake, due to the air compression
when the pile falls down. Thus we find the basic equation governing the
problem :%

\begin{equation}
\frac{K\Delta P}{Dh_{C}}=\rho\left(  \frac{1}{18}\frac{C}{1-C}p\sin
\theta\right)  gh_{C} \label{basic}%
\end{equation}

Written in this form, Eq. (\ref{basic}) can be seen as describing the balance
between two antagonistic pressures :

\begin{itemize}
\item  An ''hydrostatic'' pressure $P_{g}$ which accounts for the screening
effect of the avalanche properties of the powder%

\begin{equation}
P_{g}=\rho^{\ast}gh_{C}%
\end{equation}

where $\rho^{\ast}=\rho\left(  \frac{1}{18}\frac{C}{1-C}p\sin\theta\right)  $
is the normalized density of the particles sitting near the apices and
participating to the avalanches.

\item  The equivalent of a Laplace-Young pressure, $P_{l}$ (describing the
pressure difference at the interface of two liquids) which can be written%

\begin{equation}
P_{l}=\frac{K\Delta P}{Dh_{C}}=\gamma^{\ast}\left(  \frac{2}{h_{C}}\right)
\label{pressure}%
\end{equation}

where $\gamma^{\ast}$ plays the role of a surface tension and is defined by%

\begin{equation}
\gamma^{\ast}=\frac{K\Delta P}{2D} \label{gamma}%
\end{equation}
\end{itemize}

In brief, Eq. (\ref{basic}) describes the equilibrium of the analogue of a
wetting liquid droplet\cite{degennes85} on an horizontal plate.\ Thus, we
assimilate a conical powder pile with a half spherical wetting material of
height $h_{C}$ and curvature $2/h_{C}.$ This ersatz displays a surface tension
(or capillary forces) given by equation \ref{gamma}.\ This analogue to a
surface tension can be seen as resulting from the convective forces
(Fig.\ref{fig1}) which drag powder particles from the surrounding surface and
subsequently inject them into the powder pile. Therefore, the equivalent
surface tension of the powder pile has a purely dynamical origin since it
results from the convective forces related to the volcano effect. From
Eq.\ \ref{basic}, we get $h_{C}$ from the following relationship%

\begin{equation}
h_{C}\simeq\left(  \frac{K\Delta P}{D}\frac{1}{\rho^{\ast}g}\right)
^{\frac{1}{2}}=\left(  \frac{2\gamma^{\ast}}{\rho^{\ast}g}\right)  ^{\frac
{1}{2}} \label{capillary}%
\end{equation}

Going on with the analogy to wetting liquids\cite{degennes85}, we can also
define the usual capillary length $\lambda$ equating the hydrostatic pressure
and the Laplace-Young pressure so that $\lambda=\left(  \gamma^{\ast}%
/\rho^{\ast}g\right)  ^{\frac{1}{2}}=h_{C}/\sqrt{2}$ and a related Bond number
$Bo=\left(  \rho^{\ast}gh_{C}^{2}/\gamma^{\ast}\right)  $

Now, using Eq. (\ref{lambda}), we find%

\begin{equation}
\Lambda=\sqrt{\frac{\pi}{3e\tan^{2}\theta}}\left(  \frac{K\Delta P}{D}\frac
{1}{\rho^{\ast}g}\right)  ^{\frac{3}{4}}=\sqrt{\frac{\pi}{3e\tan^{2}\theta}%
}\left(  \frac{2\gamma^{\ast}}{\rho^{\ast}g}\right)  ^{\frac{3}{4}}
\label{lambda1}%
\end{equation}

Here a numerical estimation of the involved parameters is imperative. We
calculate an approximate value for the surface tension $\gamma^{\ast}$starting
from Eq.\ (\ref{lambda1}) using typical values for $\Lambda(5mm),$ e$(20\mu
m)$ and $\rho^{\ast}$ obtained for $C=5\%.$ We get $\gamma^{\ast}%
\simeq2.\,3\ast10^{-5}Nm^{-1}$ which means that this constant is about $3000$
times smaller than the surface tension of pure water. As expected, $\lambda$
and $h_{C}$ are in the order of 1mm. Moreover, starting from
Eq.\ (\ref{pressure}) we can get an estimated value for the pressure
difference between the altitude $h_{C}$ and the base. First, we consider that
the permeability of the granular material is a fraction of the cross sectional
area of a single particle. Thus, we get $\Delta P$ in the order of $3$ Pascal.
This quantity should be a fraction of the maximum possible air pressure due to
the total weight of the powder pile leaning on the basis surface $S$. \ This
maximum air pressure is found to be about $10\;$Pascal which therefore stands
as a correct order of magnitude.

Table \ref{table1} summarizes the analogy between the basic equations
governing the powder heap equilibrium and the equations governing the
equilibrium of liquid droplets.%

%TCIMACRO{\TeXButton{B}{\begin{table}[tbp] \centering}}%
%BeginExpansion
\begin{table}[tbp] \centering
%EndExpansion%
\begin{tabular}
[c]{|l|l|l|l|}\hline
\textbf{Wetting liquid} & \textbf{Eq.} & \textbf{Blown powder heap} &
\textbf{Eq.}\\\hline
Surface tension & $\gamma=\frac{dF}{dl}$ & Convective forces & $\gamma^{\ast
}=\frac{K\Delta P}{2D}$\\
Droplet radius & $R$ & heap height & $h_{C}$\\
Laplace-Young law & $\Delta P=\frac{2\gamma}{R}$ & Eq. \ref{pressure} &
$\Delta P^{\ast}=\frac{2\gamma^{\ast}}{h_{C}}$\\
droplet equilibrium & $\frac{2\gamma}{R}=\rho gR$ & blown heap equilibrium &
$\frac{2\gamma^{\ast}}{h_{C}}=\rho^{\ast}gh_{C}$\\\hline
\end{tabular}
\caption{Basic equations for a wetting liquid and a  blown powder\label
{table1}}%
%TCIMACRO{\TeXButton{E}{\end{table}}}%
%BeginExpansion
\end{table}%
%EndExpansion

\begin{figure}
[h]
\begin{center}
\includegraphics[
height=1.7928in,
width=3.4342in
]%
{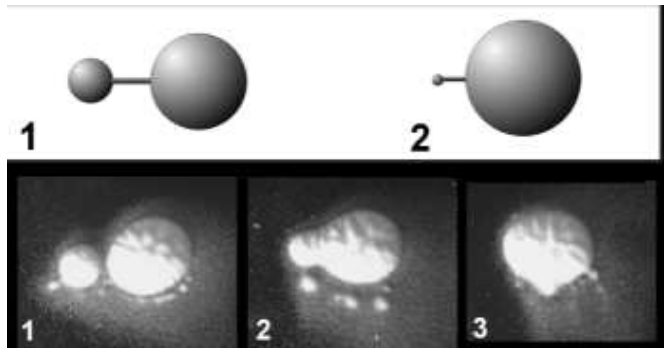}%
\caption{Above, sketch of a basic experiment showing that the inner gas
pressure is larger in a smaller droplet. When two bubbles are connected by a
small pipe, the small bubble is sucked into the largest one.\ \ Below, the
bird eye view of an experiment showing the fusion mechanism among tapped
powder heaps.\ The smaller heaps are sucked into the largest heap in agrement
with eq. \ref{basic}}%
\label{fig4}%
\end{center}
\end{figure}

Now, starting from this analogy and using e.g. Eq. (\ref{basic}), we can
transcribe the classical demonstration of the Rayleigh-Taylor instability for
wetting liquids.\ The standard analysis consists in examining the evolution of
an infinitesimal sinusoidal distortion of the initially flat surface.\ Note
that the basic calculation for liquids (found in text-books) leads to a
wavelength dependence $\Lambda\propto\left(  \gamma/\rho g\right)  ^{\frac
{1}{2}}.$ Here, the distortion is by no means infinitesimal.\ We rather
introduced the volume conservation condition (Eq. (\ref{volume})) which leads
to $\Lambda\propto\left(  \gamma^{\ast}/\rho^{\ast}g\right)  ^{\frac{3}{4}}%
.$\ But except for this difference, the underlying phenomenology of the blown
powder mimics the standard Rayleigh-Taylor instability.

\ Still proceeding with the analogy of the inner pressure within a powder heap
given \ by Eq. (\ref{pressure}) which mimics the Laplace-Young law, we predict
that if two powder heaps of unequal sizes are sitting next to each other, the
smaller one would be sucked into the larger one just as this occurs between
two communicating bubbles.\ The experimental result fulfills this assertion as
can be seen in Fig.\ref{fig4}.%

%TCIMACRO{\FRAME{fhFU}{3.4342in}{1.7928in}{0pt}{\Qcb{Above, sketch of a basic
%experiment showing that the inner gas pressure is larger in a smaller droplet.
%When two bubbles are connected by a small pipe, the small bubble is sucked
%into the largest one.\ \ Below, the bird eye view of an experiment showing the
%fusion mechanism among tapped powder heaps.\ The smaller heaps are sucked into
%the largest heap in agrement with eq. \ref{basic}}}{\Qlb{fig4}}{fig4.eps}%
%width 3.4342in;  height 1.7928in;  depth 0pt;  original-width 9.5986in;
%original-height 4.99in;  cropleft "0";  croptop "1";  cropright "1";
%cropbottom "0";  filename 'fig4.eps';file-properties "XNPEU";}}}%
%BeginExpansion
%EndExpansion

\begin{figure}
[h]
\begin{center}
\includegraphics[
height=2.2779in,
width=2.8669in
]%
{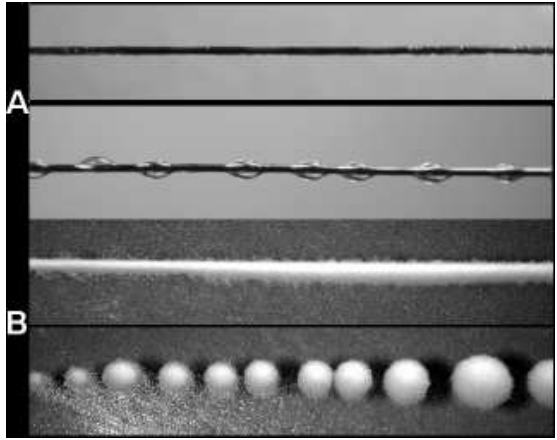}%
\caption{Fig. A\ is a snapshot of the spontaneous droplet formation when a
thin rod is covered with a wetting film of maple syprup (above). The liquid
films splits into small droplets (below) Fig. B is a bird eye view of an
seemingly analogous behaviour using a fine powder.\ Starting from a line of
fine powder (above), the support is tapped, the thin line splits into quasi
regularly spaced powder piles (below). }%
\label{fig5}%
\end{center}
\end{figure}

Furthermore, it is well known that a thin film of a wetting liquid spread over
a cylindrical rod is intrinsically unstable. Due the so-called Rayleigh
instability, such a cylindrical film splits into a collection of approximately
equally spaced droplets because of the curvature of the supporting surface.
Here, we find a seemingly identical behavior starting from a single line of
powder spread over a horizontal plate (Fig.\ref{fig5}).\ %

%TCIMACRO{\FRAME{fhFU}{2.8669in}{2.2779in}{0pt}{\Qcb{Fig. A\ is a snapshot of
%the spontaneous droplet formation when a thin rod is covered with a wetting
%film of maple syprup (above). The liquid films splits into small droplets
%(below) Fig. B is a bird eye view of an seemingly analogous behaviour using a
%fine powder.\ Starting from a line of fine powder (above), the support is
%tapped, the thin line splits into quasi regularly spaced powder piles (below).
%}}{\Qlb{fig5}}{fig5.eps}{\special{ language "Scientific Word";
%type "GRAPHIC";  maintain-aspect-ratio TRUE;  display "USEDEF";
%valid_file "F";  width 2.8669in;  height 2.2779in;  depth 0pt;
%original-width 8.8894in;  original-height 7.05in;  cropleft "0";
%filename 'fig5.eps';file-properties "XNPEU";}}}%
%BeginExpansion
%EndExpansion

However, the analogy is misleading here because the curvature of the support
does not play any role in this experiment.\ We should rather consider this
result as a 1D\ version of the Rayleigh-Taylor instability in fine powders.

At this point, it is meaningful that the ambient fluid viscosity $\eta$
cancels out in Eq. (\ref{basic}) just as in the classical analysis of the
steady state of the Rayleigh Taylor instability. It is also certainly not
fortuitous that the similarity between the leading equation for wetting fluids
and blown fine powders comes from the formal analogy between the Darcy's and
Laplace-Young law. \ In connection with that, we note that the equivalent
surface tension $\gamma^{\ast}$ is in the order of $D\Delta P.$ The
corresponding energy due to a small surface change $dS$ reads $\delta W\sim$
$\gamma^{\ast}dS\sim D^{3}\Delta P\sim V_{p}\Delta P$ where $V_{p}$ is the
volume of a single pore of the granular cake.\ This merely illustrates the
fact that the Darcy's law expresses the Poiseuille law across a single pore of
the porous cake.

Even if it has the merit to establish a connection between the (yet unknown)
description of blown powder properties and the (already known) wetting liquids
behavior, our simple theoretical explanation certainly lays itself open to
several criticisms.\ In particular, it does not convey any information
regarding the development of the surface instability.\ Such an analysis would
involve the introduction of a sort of powder viscosity\cite{thoroddsen01}
which is not considered in the present model dealing with the steady state of
the process. A time resolved scrutiny of the pattern growth would probably
convey information about this question.\ We postpone the description of this
study to a forthcoming paper.

I am grateful to R.\ Jacobs, E.\ Raphael, I.\ Aronson and the granular group
in Jussieu for stimulating discussions.

\end{document}